\documentclass[%
 reprint,
 amsmath,amssymb,
 aps,
pre,
]{revtex4-2}

\usepackage{graphicx}
\usepackage{dcolumn}
\usepackage{bm}
\usepackage{pbox}
\usepackage{multirow}
\usepackage{svg}

\newcommand{\Ordo}[1]{\mathcal{O}\left(#1\right)}

\begin{document}

\preprint{APS/123-QED}

\title{Classical fractional time series from quantum XXZ spin chains}



\author{Zoltán Udvarnoki\textsuperscript{1}}
\email{zoltan.andras.udvarnoki@ttk.elte.hu}
\author{Gábor Fáth\textsuperscript{1,2}}
\author{Miklós Werner\textsuperscript{3}}
\author{Örs Legeza\textsuperscript{3,4,5}}

\affiliation{\textsuperscript{1}Department of Physics of Complex Systems, Eötvös Loránd University, Budapest, Hungary}
\affiliation{\textsuperscript{2}MPP\&E Capital, London, UK}
\affiliation{\textsuperscript{3}Wigner Research Centre for Physics, Budapest, Hungary}
\affiliation{\textsuperscript{4}Institute for Advanced Study,Technical University of Munich, Lichtenbergstrasse 2a, 85748 Garching, Germany
}
\affiliation{\textsuperscript{5}Parmenides Stiftung, Hindenburgstr. 15, 82343, Pöcking, Germany
}

\date{\today}

\begin{abstract}
Entangled quantum mechanical states in one dimension can be used to represent and simulate classical stochastic processes with nontrivial statistical properties. Long-range quantum correlations translate into fractional processes with their asymptotic Hurst exponents characterizing roughness and persistence. We explore this analogy in the case of the spin-1/2 XXZ chain and investigate properties of four different classical two-state processes that this quantum system can generate. These processes show fractional characteristics with varying Hurst exponents. We argue that the continuous quantum symmetries such as U(1) or SU(2) of the XXZ chain give rise to $H=0$ with logarithmic scaling. Processes generated without these symmetries can produce $H \geq0.5$ but likely not $H < 0.5$ unless the dominant term responsible for $H=0.5$ gets canceled. This does not seem to happen for the XXZ model. We use standard quantum methods, including MERA and TEBD, to numerically substantiate our findings.

\end{abstract}

\maketitle


\section{Introduction}
Fractional processes are stochastic processes with long-range dependence, asymptotic self-similarity, and scale-invariant behavior. They have been applied to many fields, such as economics \cite{Lee2022GeneralizedDistribution}, physics and geophysics \cite{Scipioni2008CharacterizationPlasmas, Molz1997FractionalExtensions}, computer science \cite{Beran1995Long-rangeTraffic, Leland1994OnVersion}, financial modeling \cite{Jamdee2005LongValuation, Tapiero2016TheFinance}, especially in stochastic volatility models \cite{ Livieri2018RoughPrices, Gatheral2018, David2021FractionalSeries}. Fractional processes are represented by scaling exponents, one or many, which determine the asymptotic scaling of their moments, such as their variance. Representing and sampling fractional processes is not straightforward and are an active field of research \cite{Wong2024SimulationApproach}. 

Power-law scaling of moments arises from slowly decaying correlations of the process increments. Certain quantum mechanical systems, such as quantum spin chains, show similar correlation functions at their critical point. This creates an interesting analogy between the two systems, which may be used to represent and generate fractional time series. The quantum analogy potentially opens a way to sample fractional processes directly from carefully engineered physical systems with quantum measurement, using quantum computers, or via quantum-inspired methods on classical hardware.

In this study, we explore this analogy through the lens of the Heisenberg XXZ quantum spin chain, which is a well-studied, standard model of one-dimensional strongly interacting discrete quantum degrees of freedom with a rich phase diagram containing an asymptotically self-similar (critical) region with power-law correlations. We demonstrate how the XXZ model can be used to generate fractional series of two states. Having parameter-dependent critical exponents makes the model a potential candidate to generate fractional processes with Hurst exponents both below and above 0.5. Note, however, that the Heisenberg chain is not self-similar in the strict sense, self-similarity only sets in asymptotically. This brings in a dependence on the specific short-range properties of the model. As we will argue, this interferes with and modifies the naively expected scaling behavior for the generated classical process. 
The analogy between time series and quantum spin chains was already suggested \cite{Udvarnoki2025}, while MPS has been used extensively in recent studies for machine learning tasks such as generative modeling \cite{Han2018UnsupervisedStates, Stokes2019ProbabilisticStates, Liu2023, Kobayashi2024} and classification \cite{Stoudenmire2016, Harvey2025, Mossi2025}. 

The paper is structured as follows: in Sections \ref{sec:frac_time_series} and \ref{sec:xxz_model}, background information and known results about fractional time series and the XXZ model are introduced, then in Section \ref{sec:var_scaling_analytic}, analytical considerations regarding the scaling of the different processes' variance are discussed, while Section \ref{sec:numerical_results}, the results of the numerical calculations are presented. Section \ref{sec:simulation} describes sampling techniques for fractional processes and tensor networks. Finally, Section \ref{sec:conclusion} summarizes our conclusions.

\section{Fractional time series}
\label{sec:frac_time_series}

\subsection{Fractional Gaussian noise and Brownian motion}

The archetype of fractional processes is the fractional Brownian motion (fBM).\cite{Mandelbrot1968FractionalApplications} With our forthcoming line of reasoning in mind, we define it here in discrete time through its increments. We start with fractional Gaussian noise (fGN) $X_t$, which is a stationary Gaussian process with auto-covariance
\begin{eqnarray}
   C(n) &=& \langle X_{t} X_{t+n} \rangle = \nonumber \\
   && \frac{\sigma^2}{2}\left( |n+1|^{2H} -2|n|^{2H} + |n-1|^{2H} \right),
   \label{eq:fGN-covar}
\end{eqnarray}
where $n\ge 0$ is an integer time lag, $\sigma^2 = \langle X_{t} X_{t} \rangle$ is the constant variance of fGN, and $0<H<1$ is its Hurst exponent. For large time distances, $C(n)$ behaves like a power law:
\begin{equation}
C(n) \sim  \sigma^2 H(2H-1)n^{2H-2}, \quad \mathrm{for}\, n \gg 1.
\end{equation}

Fractional Brownian motion (fBM) is defined as the cumulative process of fGN increments
\begin{equation}
    B_t = \sum_{i=1}^t X_i.
\end{equation}
As a result of the covariance structure in Eq.\ (\ref{eq:fGN-covar}), fBM has a time-dependent variance that behaves exactly as a power law with exponent $2H$ (see Section \ref{sec:fbm_var} for a derivation of how the short-term properties of $C(n)$ play a role in this):
\begin{equation}
\text{Var}(B_t) = \sigma^2 t^{2H}.
\end{equation}
Fractional Brownian motion is exactly self-similar (not just asymptotically) in terms of its variance.

\subsection{Generalized Bernoulli process and fractional binomial process}
Discrete state processes are better suited for the analogy with finite-level quantum systems.
The generalized Bernoulli process (gBP) proposed by Lee \cite{Lee2020} is a discretized version of the fractional Gaussian noise. The process built up of these increments is called the fractional binomial process (fBP).

The generalized Bernoulli process is a two-state stochastic process with $Prob(1)=p$, $Prob(0)=1-p$, defined through its auto-covariance function:
\begin{equation}
\label{eq:GBP-cov}
\langle X_t X_{t+n} \rangle  = c n^{2h-2}, \quad n >= 1.
\end{equation}
with $c$ and $h$ constants. This mimics the asymptotic form of the fGN covariance in Eq. \eqref{eq:fGN-covar}, but not its specific short-term features.

The variance of the fBP for large distances turns out to be \cite{Lee2020}:
\begin{equation}
\label{eq:payoff}
  Var(B_t)  \propto 
  \begin{cases}
  t \quad &h \in (0,0.5)\\ 
  t\ln t \quad &h=0.5 \\
  t^{2h} \quad &h \in (0.5,1)
  \end{cases}
\end{equation}
Note that, contrary to the fBM, the fraction Binomial process does not produce rough scaling for $h<0.5$. The variance cannot grow slower than linear, its Hurst exponent is $H=1/2$. Sublinear scaling is suppressed by an $H=1/2$ scaling operator, which is not canceled as in the case of the fBM without an appropriately engineered short-term structure.

\section{Heisenberg XXZ 1/2-spin chain}
\label{sec:xxz_model}

The Heisenberg XXZ 1/2-spin chain with anisotropy parameter $\Delta$ is defined by the Hamiltonian:
\begin{equation}
    H = J\sum_{j=1}^L \left[ \left(S^x_j S^x_{j+1} + S^y_j S^y_{j+1} \right) + \Delta S^z_j S^z_{j+1} \right],
\end{equation}
where $S^a_j$ is the spin-$\frac{1}{2}$ projection operator in direction $a$ acting on site $j$, $J$ is the coupling, chosen to be $1$ and $\Delta$ is the anisotropy parameter.

The system is in a gapless phase if $|\Delta| < 1$ and experiences long-range correlations in its paramagnetic ground state. Asymptotic results are known for the correlation of spin-spin operators from field theoretical calculations. Exact results are hard to calculate but are known for small distances \cite{Shiroishi2005ExactChain, Sato2005CorrelationFunction}. For $\Delta = 0$ the $\langle S^z_i S^z_{i+l} \rangle$ auto-covariance takes a simple form and the $\langle S^x_i S^x_{i+l} \rangle$ is also calculable in closed form \cite{Kitanine2002CorrelationRepresentations}. 

The leading term in the critical region is an algebraically decaying correlation for all directions \cite{Collura2017}, similar to the auto-covariance of the GBP (Eq. \eqref{eq:GBP-cov}). The dominating terms are:
\begin{eqnarray}
C_x(n) &=& \langle \mathrm{GS}|S_{j}^{x}S_{j+n}^{x}|\mathrm{GS} \rangle = (-1)^n \frac{\tilde{A}}{n^\eta} + \frac{A}{n^{\eta+1/\eta}}  + \ldots \nonumber \\
C_z(n) &=& \langle \mathrm{GS}|S_{j}^{z}S_{j+n}^{z}|\mathrm{GS} \rangle = (-1)^n \frac{\tilde{A}}{n^{1/\eta}} -\frac{A}{n^2} + \ldots, \nonumber \\
\label{eq:spin_cov}
\end{eqnarray}
where 
\begin{equation}
\Delta = -\cos(\pi\eta) \label{eq:delcos} 
\end{equation}
connects scaling exponents to spin chain anisotropy. 

The asymptotic behavior is a sum of a monotonically decaying and an oscillating term. Hence, it is customary to define and study two spin aggregation functions, the "uniform" and the "staggered" domain magnetizations over length $l$. With longitudinal and transverse directions wrt the direction of anisotropy, we have four interesting quantities to study:

\begin{eqnarray}
M_{z}(l) &= \sum_{k=1}^l S^{z}_k,\qquad N_z(l) &= \sum_{k=1}^l (-1)^k S^{z}_k\nonumber\\
M_{x}(l) &= \sum_{k=1}^l S^{x}_k,\qquad N_x(l) &= \sum_{k=1}^l (-1)^k S^{x}_k.
\end{eqnarray}

\begin{table*}[t]
\centering
\begin{tabular}{|l|l|l|}
\hline
 & Classical stochastic example & Quantum spin chain \\
\hline
Increment process & GBP or fGN & Spin-1/2 quantum chain \\
Integrated process & fBm of fBP & Spin domain magnetization \\
Correlations & Power law (in time) & Power law (in space) \\
Fractal characteristics & Hurst exponent & Correlation exponents \\
Probability measure & Conditional Bernoulli/normal & Quantum ground state implied \\
Sample & Process trajectory & Quantum measurement \\ 
Sampling algorithm & \pbox{20cm}{\vspace{.5\baselineskip}Cholesky, Circulant, Kernel \\ or by the definition\vspace{.5\baselineskip}} & \pbox{20cm}{\vspace{.5\baselineskip}TN (MPS, MERA) sampling\\ or physical measurement \vspace{.5\baselineskip}} \\
\hline
\end{tabular}
\caption{The details of the analogy between stochastic processes and quantum spin chains.}
\label{tab:analogy}
\end{table*}

Spin chains provide a natural analogy with fractional stochastic processes based on their stationary, zero-mean increments and asymptotic scaling of the autocovariance of the increments (see Table \ref{tab:analogy}). In Ref \cite{Udvarnoki2025} we showed how to leverage this analogy to generate classical time series from quantum states such as the ground state of the XXZ model. 
The fractal properties of the time series generated will reflect the statistical properties of the magnetic domains. Hence, in the next section, we will investigate these in detail.

\section{Scaling of the variance of magnetic domains}
\label{sec:var_scaling_analytic}

In general, we can calculate the variance of any cumulative process using the increments' covariance function.
With $X_t$ being the increment process, we define the cumulative process as $B_t = \sum_{i=1} ^{t} X_i$. When the increment process has zero mean $\langle B_t \rangle = 0$, the variance of the cumulative process can be expressed as:
\begin{eqnarray}
\label{eq:gen_var}
    \mathrm{Var}(B_t) = \langle B_t^2 \rangle &=& \left\langle\left(\sum_{i=1} ^t X_i\right)\left(\sum_{i=1} ^{t} X_i\right)\right\rangle \nonumber \\ &=&\sum_{i,j=1}^t\langle X_i X_j \rangle \nonumber \\ &=&  \sum_{i=1}^t \left\langle X_i^2 \right \rangle + 2 \sum_{i=1}^t\sum_{j=1}^{t-i} C(j) \nonumber  \\
    &=&  \sigma^2 t + 2 \sum_{i=1}^t\sum_{j=1}^{t-i} C(j),
\end{eqnarray}
where we have used that increments are stationary with $\left\langle X_i^2 \right\rangle = \sigma^2$. The first, linear term is always present irrespectively of the autocovariance structure and in itself alone would give rise to $H=0.5$ scaling.

Asymptotic scaling of the variance can be derived from the scaling of the covariance function. To see this, look at the double sum as a double integral, which is a valid approximation in the large domain length limit. If there is a nonzero constant term coming from the first integral, it will be integrated again, leading to a term proportional to $l$. This $\Ordo l$ term is irrelevant when $C(l)\sim l^{-\alpha}, \quad \alpha<1$, i.e., decays slower than $1/l$, because then the dominant contribution from the double integral is $\Ordo {l^{2-\alpha}}$ which increases faster than linear. This superlinear scaling leads to a Hurst exponent $H > 0.5$ as expected.

On the other hand, when $C(l)\sim l^{-\alpha}$ with $\quad \alpha<1$, i.e., covariance decays faster than $1/l$, the double integral scales as $\Ordo {l^{2-\alpha}} + \Ordo {l}$ in which $\Ordo {l}$ dominates unless the constant term of the first integral vanishes for any reason. Without this the Hurst exponent is $H=0.5$.

\subsection{Fractional Brownian motion}
\label{sec:fbm_var}
It is illuminating to evaluate the double sum term exactly for fractional Brownian motion and see how it interacts with the linear term. With the covariance function in Eq.\ (\ref{eq:fGN-covar}), the inner sum of the second term is a telescopic series:
\begin{eqnarray}
    \sum_{j=1}^{k} C(j) &=& \frac{\sigma^2}{2} \left(0 - 2  + 1 + k^{2H} -2k^{2H} + (k+1)^{2H} \right) \nonumber \\&=& -\frac{\sigma^2}{2}(1 + k^{2H} -(k+1)^{2H}).
\end{eqnarray}
Substituting back we get a telescopic series again in the outer sum. The whole expression becomes
\begin{eqnarray}
\mathrm{Var}(B_t)  &=& \sigma^2t - \sigma^2\left( t + \sum_{i=0}^{t-1} [i^{2H}- (i+1)^{2H}]\right) \nonumber \\ &=& \sigma^2 \sum_{i=0}^{t-1} [(i+1)^{2H} - i^{2H}] = \sigma^2 t^{2H}.
\end{eqnarray}
Notice that the linear term gets canceled exactly. This "miraculous" cancellation is the consequence of the particular small-distance auto-covariance structure of the fractional Gaussian noise process. What remains for $\mathrm{Var}(B_t)$ is a single power-law term whose exponent is $2H$ and thus behaves sub- or super-linearly below or above $H=0.5$.

\subsection{Generalized Bernoulli process}

In this case the linear term does not cancel as was already found by J.\ Lee.\cite{Lee2020}. The variance becomes dominated by this linear scaling term for $h<0.5$, and this suppresses sub-linear scaling. For $h > 0.5$, super-linear scaling arises as expected, and the Hurst exponent is $H=h$.

\subsection{Heisenberg spin chain}
\label{sec:spinchain_var}

For a Gaussian process, like the fGn or the fBm, the expectation value and the (auto)covariance function are sufficient to define the process. This is not true for non-Gaussian processes, like the fBP or gBP. Many processes with the same expectation value and covariance structure can be constructed.

A major property of fractional processes is the power-law scaling of the variance with the Hurst exponent. In this study, we will only concentrate on the variance (second central moment) of a given process when analyzing its fractional and self-similar nature, arguing that this can be easily obtained for real-world time series.

Without knowing the exact short-distance form of the auto-covariance, it is impossible to calculate the variance of the magnetization. Even the asymptotic behavior can depend on the unknown short-distance behavior.

It is important to see the contribution of the staggered term of the covariance to the uniform magnetization and vice-versa for any real $p$ (for derivation see Appendix \ref{appendix:altsum}):
\begin{eqnarray}
    \sum_{k=1}^n (-1)^k k^p &=& \frac{(-1)^n}{4}(2n^p + pn^{p-1}) + C_1 \nonumber \\  
    & & + O(l^{p-3})\\
    \sum_{n=1}^l\!\sum_{k=1}^n (-1)^k k^p &=& \frac{(-1)^l}{4}l^p\! + C_1 l + C_2 + O(l^{p\!-\!1})
\end{eqnarray}

In summary, if the autocovariance of the increments has the long-distance asymptotic of:
\begin{equation}
    C(l)  = A l^{-\zeta}+ (-1)^l \tilde{A} l^{-\xi}, 
\end{equation}

then the long-distance asymptotics of the variance of the uniform cumulative process are:
\begin{equation}
    \langle M^2(l) \rangle = Bl^{-\zeta + 2} + \tilde{B} l^{-\xi} + Cl
\end{equation}
and
\begin{equation}
    \langle N^2(l) \rangle = Dl^{-\zeta} + \tilde{D} l^{-\xi+ 2} + Fl,
    \label{eq:Nvar}
\end{equation}
if $\zeta, \xi \notin \{1,2\}$. Only if the constants $C$ or $F$ are $0$, can we expect a behavior that is similar to the fBm with $H<0.5$. These calculations can be applied to the Heisenberg spin chain to predict the asymptotic behavior of the domain magnetization's variance. 
The resulting exponents can be seen in Fig. \ref{fig:var_exponents} with a dashed line if $C$ or $F$ is $0$ and with a solid line if not $0$.

There are two cases where we can make stronger statements: a) when $\Delta = 0$ and b) when there is a conserved quantum number behind the generating quantum system.

\subsection{Case of $\Delta = 0$}

At the XX point, $\Delta=0$, the exact form of the covariance function is known exactly \cite{Kitanine2002CorrelationRepresentations}: 
\begin{widetext}
\begin{equation}
    C_z(n) = \frac{1}{2\pi^2 n^2} \left[(-)^n -1\right], \quad n\neq 0.
    \label{eq:autocov_z}
\end{equation}
and 
\begin{equation}
\label{eq:autocov_x}
    C_x(n) =  \frac{(-1)^m}{4} \prod_{k=1}^{[\frac{m}{2}]} \frac{\Gamma^2(k)}{\Gamma(k - \frac{1}{2})\Gamma(k + \frac{1}{2})} \prod_{k=1}^{[\frac{m+1}{2}]} \frac{\Gamma^2(k)}{\Gamma(k - \frac{1}{2})\Gamma(k + \frac{1}{2})}.
\end{equation}
\end{widetext}
Using the above formulae, the variance of the uniform and staggered domain magnetization in the $z$ direction can be calculated analytically. The calculation details are presented in Appendix \ref{sec:exact_var}, we only state the end results here.
For the uniform magnetization we find
\begin{eqnarray}
\langle M^2(l) \rangle &=& \frac{1}{4}l + 2 \sum_{i=1}^{l-1} \sum_{j=i+1}^l C(j-i) \\
&=& 2\ln l + \left[\frac{\pi^2}{4}+2(\gamma+1)\right] -\frac{1}{l}+\frac{6}{l^2} + O(\frac{1}{l^3}), \nonumber
\end{eqnarray}
and the staggered magnetization reads
\begin{eqnarray}
\langle N^2(l) \rangle &=& \frac{1}{4}l + 2 \sum_{i=1}^{l-1} \sum_{j=i+1}^l (-)^{j-i} C(j-i) \\
&=& \frac{l}{2} -2\ln l - \left[\frac{\pi^2}{4}+2(\gamma+1)\right] +\frac{1}{l} -\frac{6}{l^2} + O(\frac{1}{l^3}). \nonumber
\end{eqnarray}
We refer to the Appendix to see how the linear term vanishes exactly in the uniform case.

\begin{figure}[h]
    \centering
    \includegraphics[width=0.9\linewidth]{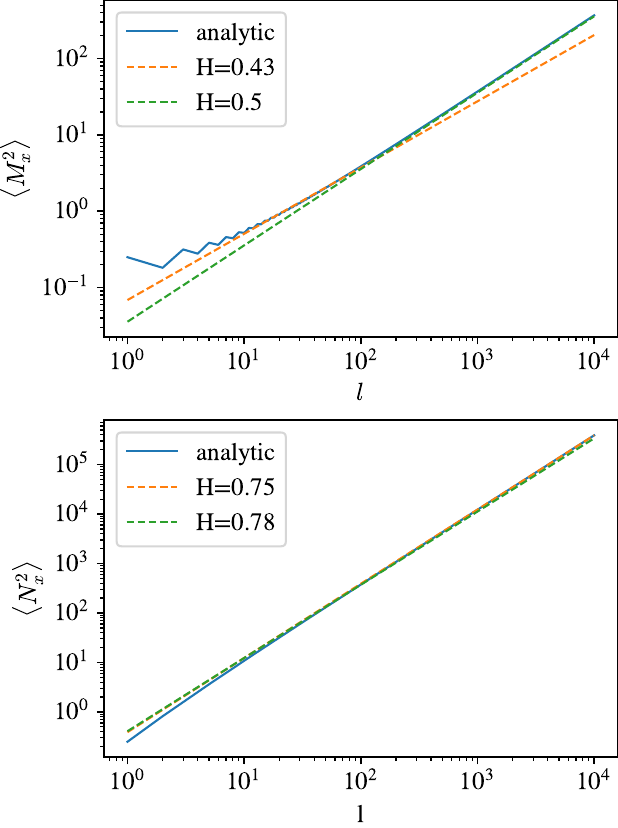}
    \caption{The variance of the uniform domain magnetization in the $x$ direction using the analytical formula. The dashed lines show the behavior with different Hurst exponents for small and large domain sizes.}
    \label{fig:XX}
\end{figure}

For the $x$ direction, we cannot derive results analytically, but we can use the exact form of the covariance function and do the summation numerically for long distances. Results are shown in Fig.\ \ref{fig:XX}. Behavior of the uniform magnetization is consistent asymptotically with $H=1/2$, but we need long chains of hundreds of spins to see this with precision. This means that the linear term of the double summation does not vanish and dominates the behavior. On shorter distances, the empirically fitted exponent can appear misleadingly smaller. 

The staggered magnetization can be fitted well with an asymptotic power law with $H\sim 0.75$. This is consistent with the exponent $\xi=1/2$ of the covariance function through Eqs.\ (\ref{eq:delcos}) and (\ref{eq:Nvar}). Since this is superlinear, the linear term does not count.

\subsection{Case of a quantum symmetry}

The $U(1)$ symmetry of the model, i.e., the invariance under the simultaneous rotation of spins around the $z$-axis, leads to the conservation of the $z$-component of the total spin, and this has important bearings on the scaling of the uniform magnetization $M_z$. Suppose, for a moment, a chain of length $L$ with periodic boundary conditions. Assume that $L$ is even, which makes the derivation simpler but has no relevance in the $L\rightarrow \infty$ limit. 

For $|\Delta| \le 1$, the gapless ground state of the model is an eigenstate of $M_z(L)$ with zero eigenvalue,
\begin{equation}
M_{z}(L) |\mathrm{GS} \rangle = \sum_{j=1}^L S^z_j |\mathrm{GS}\rangle = 0 \; .
\end{equation}
Consequently, the variance of the magnetization is also zero,
\begin{equation}
\langle \mathrm{GS} | M_{z}^2(L) |\mathrm{GS} \rangle = \sum_{i=1}^L\sum_{j=1}^L \langle \mathrm{GS} | S^{z}_i S^z_j |\mathrm{GS}\rangle = 0 \; .
\end{equation}
We now change variables in the sums,
\begin{equation}
 0 = \sum_{j=1}^L \langle (S_j^z)^2 \rangle +  2\sum_{j=1}^L \sum_{n=1}^{L/2-1} \langle S^{z}_j S^z_{j+n} \rangle \; ,
\end{equation}
then use translational invariance and take the $L\rightarrow \infty$ limit to get the simple sum rule,
\begin{equation}
 0 = \langle (S^z_j)^2 \rangle +  2\sum_{n=1}^\infty C_z(n) \; .
 \label{eq:sumrule}
\end{equation}
Note that this involves the infinite sum of the covariance function and connects it to the single-site variance.

The variance of the magnetization of a \emph{finite} segment reads as
\begin{eqnarray}
\langle M_{z}^2(l) \rangle &=& \sum_{j=1}^l \langle (S_j^z)^2 \rangle + 2 \sum_{n=1}^{l-1} \sum_{j=1}^{l-n}C_z(n) \nonumber \\
 &=& l \, \langle (S_j^z)^2 \rangle + 2\sum_{n=1}^{l-1} (l-n)C_z(n) \; .
\end{eqnarray}
Using the sum rule Eq.\ (\ref{eq:sumrule}) we get
\begin{eqnarray}
\langle M_{z}^2(l) \rangle &=& - 2\sum_{n=1}^\infty l C_z(n)
+ 2\sum_{n=1}^{l-1} (l-n)C_z(n) \; .
\end{eqnarray}
or
\begin{equation}
   \langle M_{z}^2(l) \rangle = -2 \left(\Xi^{(1)}(l) +  \Xi^{(2)}(l) \right)\; ,
\end{equation}
with
\begin{eqnarray}
   \Xi^{(1)}(l) &=& \sum_{n=1}^{l-1} n C_z(n) \; , \nonumber \\  
   \Xi^{(2)}(l) &=& l\sum_{n=l}^{\infty}C_z(n) \; .
\end{eqnarray}
Thus, we have rewritten the double sum into simple sums. It is safe to insert the asymptotic forms. This leads to an unknown constant error in $\Xi^{(1)}(l)$ which depends on the short-distance behavior of $C_z(l)$, but this will become irrelevant. The second term $\Xi^{(2)}(l)$ looks dangerous due to the $l$ prefactor, but here the sum starts only at $l$, and we can safely use the asymptotic form if $l\gg1$. By Eq.\ \eqref{eq:spin_cov} and \eqref{eq:single_alternating_sum} the two expressions read:
\begin{eqnarray}
    \Xi^{(1)}(l) &\sim& (-1)^l  \tilde{A} \frac{1}{l^{1/\eta-1}}+A \ln l + \mathrm{const} \;  \nonumber \\
    \Xi^{(2)}(l) &=& (-1)^l \tilde{A} \frac{1}{l^{1/\eta-1}} - A  l \left( \sum_{n=l}^\infty \frac{1}{n^2} \right) + \mathrm{const.} \; .
\end{eqnarray}
As $\eta < 1$, the leading term is the logarithm, i.e., for $l \gg 1$ we expect:
\begin{eqnarray}
    \langle M_z^2(l) \rangle \propto \ln l \; .
\end{eqnarray}

The $\Delta=1$ point is special in two senses: (I) not only $M_z(L)$ but also $M_x(L)$ is conserved and zero due to the full $SU(2)$ symmetry, leading to identical behavior in the two spin directions, (II) in this limit ($\eta \rightarrow 1$) the correlator $C_z(n)$ has a slightly different form~\cite{Giamarchi1989},
\begin{equation}
C_z^{\Delta=1}(n) = \frac{A}{n^2} + \frac{\tilde{A} (-1)^n}{n} \sqrt{\ln n} \; .
\end{equation}
This, however, does not affect the leading term that is still $\propto \ln l$.

\begin{table}[]
\centering
\begin{tabular}{lr|c|}
\cline{3-3}
                                                   & \multicolumn{1}{l|}{} & \multicolumn{1}{c|}{$\Delta = 0$} \\ \hline
\multicolumn{1}{|c|}{\multirow{4}{*}{Z direction}} & C(n) $\sim n^{\zeta}$        & $\zeta=-2$                            \\ \cline{2-2}
\multicolumn{1}{|c|}{} & $M^2(l)$   & $\log(l)$           \\ \cline{2-3}
\multicolumn{1}{|c|}{} & C(n) $\sim (-1)^n n^{\xi}$ & $\xi=-2$ \\ \cline{2-2}
\multicolumn{1}{|c|}{}& $N^2(l)$ &  $l$    \\ \hline \hline

\multicolumn{1}{|c|}{\multirow{4}{*}{X direction}} & C(n) $\sim n^{\zeta}$ & $\zeta = -2.5$  \\ \cline{2-2}
\multicolumn{1}{|l|}{} & $M^2(l)$     &  l              \\ \cline{2-3}
\multicolumn{1}{|c|}{} & C(n) $\sim (-1)^n n^{\xi}$ & $\xi = -0.5$  \\ \cline{2-2}
\multicolumn{1}{|l|}{} & $N^2(l)$  &  $l^{1.5}$                     \\ \hline
\end{tabular}
\caption{The asymptotic forms of the autocovariance function, the exact scaling of the domain magnetizations variance in the Z direction, and the scaling according to long-range numerical calculations in the X direction with the exact formula for $\Delta=0$.}
\label{tab:var_exact}
\end{table}

\section{Numerical results}
\label{sec:numerical_results}

We have seen above that the asymptotic scaling of (uniform and staggered) domain magnetization depends much on the short-term structure of the correlation functions and the symmetry of the quantum state. Consequently, any classical stochastic process defined from these quantum states inherits this dependency. The fractional properties that can be achieved via this quantum-classical mapping may be restricted. 

Out of the 4 processes we have defined we expect $M_z$ to scale logarithmically slowly and thus effectively represent Hurst $H=0$ due to the $U(1)$ continuous symmetry. There is no such symmetry for the other three processes except at $\Delta=1$ for $M_x$. In the general cases $M_x$, $N_z$, $N_x$ can show sub- or super linear scaling with $H < 1/2$ or $H > 1/2$ depending on $\Delta$. We see no difficulty with super-linear scaling, but sub-linear scaling requires a constant to become exactly zero in the correlation (auto-covariance) structure. Whether this condition holds cannot be determined analytically in the XXZ model. Hence we turn to numerical methods.

We calculated the ground state of the Heisenberg XXZ chains with MERA and MPS ansatzes at different anisotropy parameters. We could then calculate two-point correlations and magnetic domains of different lengths. 

The MPS was calculated using the infinite time-evolving block decimation (iTEBD) algorithm~\cite{Vidal2007iTEBD} with a bond dimension of 1,024 and exploiting one $U(1)$ symmetry, using the opeNAMPS package~\cite{Werner2020, opeNAMPS}. The scale-invariant MERA was calculated using a gradient-based method \cite{Geng2022DifferentiableNetworks} with a maximal bond dimension of 20 and two transitional layers. The number of parameters in these settings, accounting for the independent components of the tensors, is ($2\chi^2$) 2,097,152 in the case of the MPS and 334,416 for the MERA. This difference accounts for the higher accuracy of the MPS representation but is necessary because of the increased computational demands of working with the MERA ansatz.

\subsection{Autocorrelation of increments}

\begin{figure}[h]
    \centering
    \includegraphics[width=1\linewidth]{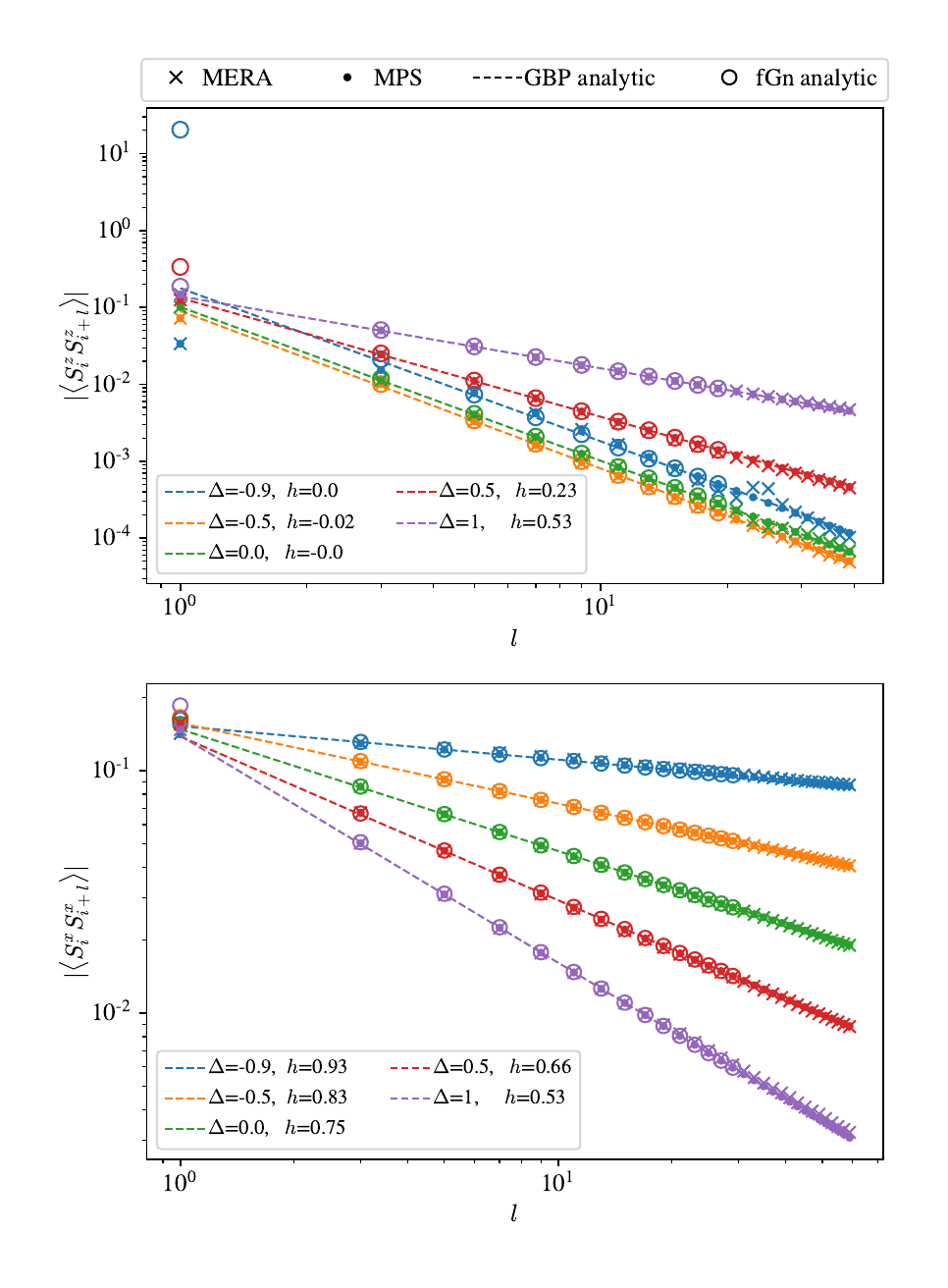}
    \caption{Auto-covariance in the X and Z direction of the spin chain calculated using MPS and MERA representation compared to the GBP and fGn analytic values. Note the difference between the fGn and the GBP for small distances. $2h-2$ is the incline of the fitted line.}
    \label{fig:corr}
\end{figure}

In Fig.\ \ref{fig:corr} we show a comparison of the short-term behavior of spin chain correlators and asymptotically equivaent fraction Brownian motion and fractional Binomial processes. Deviations are small, except for very small distances, but these can be responsible for significant qualitative alterations in domain scaling behavior.

\subsection{Scaling of the variance of the domain magnetization}
We also calculated the variance of the cumulative process (domain magnetization) using Eq. \eqref{eq:gen_var}. Looking for asymptotic scaling in the form of $l^{2H}$, we can determine the Hurst exponents $H$ numerically.

\begin{figure}[h]
    \centering
    \includegraphics[width=1\linewidth]{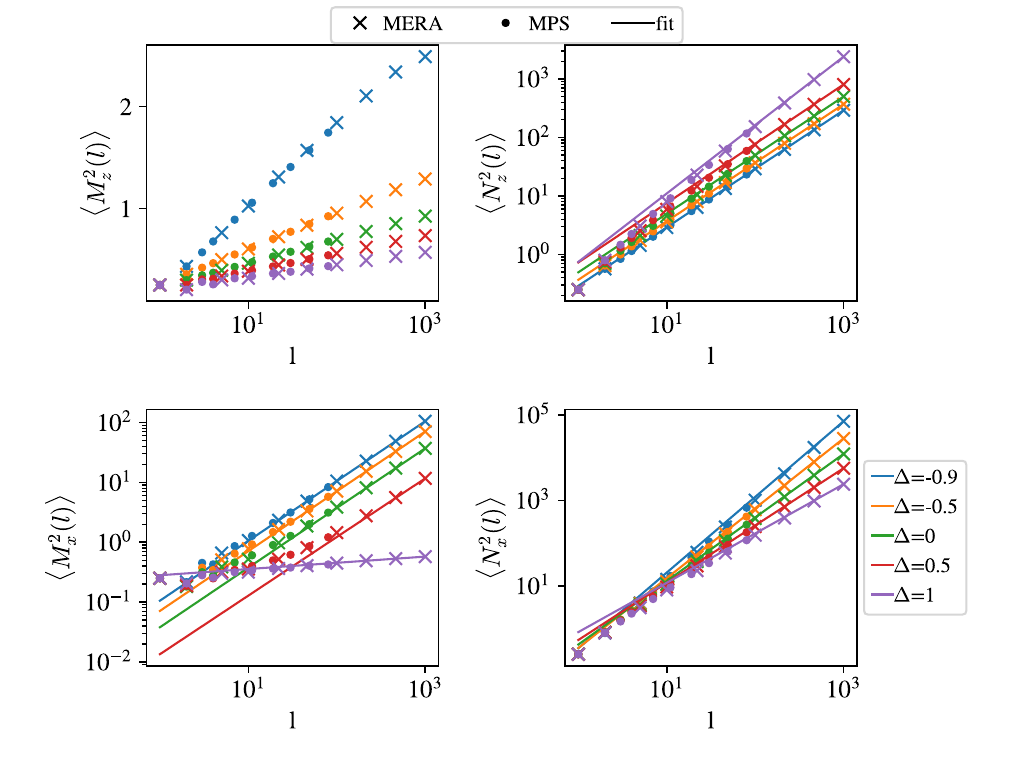}
    \caption{The variance of the uniform and staggered domain-magnetization in the $x$ and $z$ direction of the spin chain, calculated using the MPS and MERA representations for different domain lengths.}
    \label{fig:var}
\end{figure}

\begin{figure}[h]
    \centering
    \includegraphics[width=\linewidth]{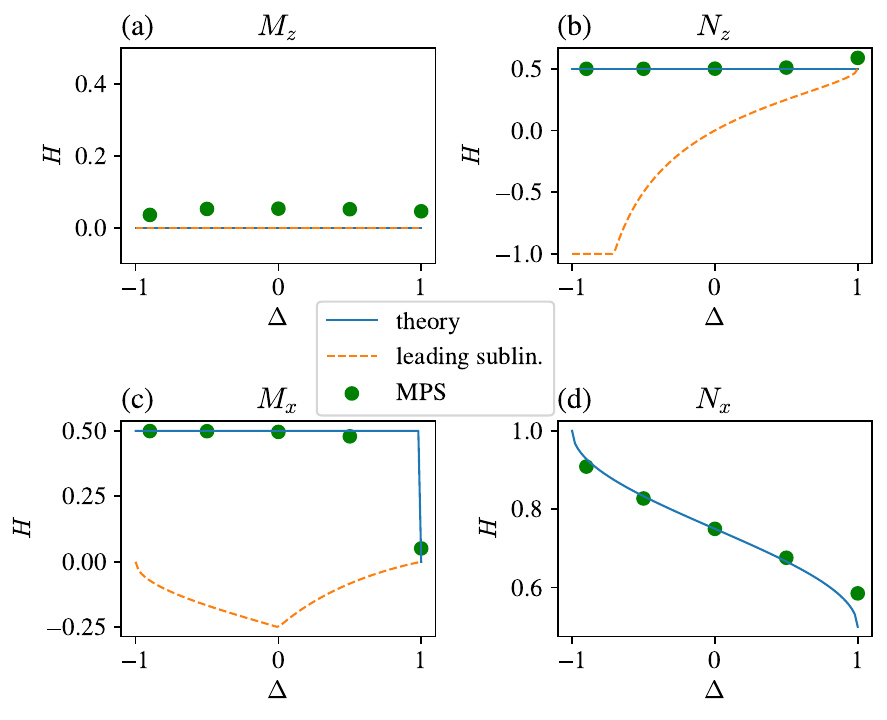}
    \caption{The exponents of the magnetization variance in different directions. Green dots indicate the fitted exponents, while the solid line represents the theoretical values. The orange dashed line indicates the exponents we would have if the linear term were canceled out.}
    \label{fig:var_exponents}
\end{figure}

We calculated the variance of domain magnetization up to a distance of 1,000 sites. Going further would accumulate numerical errors as the tensor network calculations loose precision. We will describe what can be seen up to these distances based on Fig. \ref{fig:var} and the fitted parameters in Table \ref{tab:hurst}. The behaviors are summarized in Table \ref{tab:var_scaling} and Fig. \ref{fig:var_exponents}.

\subsubsection*{Process $M_z$}

For $M_z$, the variance of the uniform magnetization scales logarithmically for all $\Delta$. This is visible by the straight lines on the lin-log plot. This is the only case where we see no power-law scaling. As discussed this is a consequence of the $U(1)$ symmetry.

\subsubsection*{Process $N_z$}


If we consider the staggered magnetization in the $z$ direction, we see a slowly increasing scaling exponent from $0.5$ to $0.58$. According to our previous calculations, this is an artifact on short distances, and all curves will converge asymptotically to the exponent $0.5$. This was shown for $\Delta=0$ analytically earlier.

\subsubsection*{Process $M_x$}

The uniform magnetization in the $x$ direction is an interesting case. Except for $\Delta=1$, we see close to linear scaling. However, for short distances, the scaling exponent tends to $0$ as $\Delta$ is increased, which would correspond to rough behavior. Only with increasing distance does the linear term of equation \eqref{eq:gen_var} take over and lead to $H=1/2$. This result is in line with Ref. \cite{Collura2017}, who found that for $|\Delta| < 1$, scaling is linear. We obtained a similar result for $\Delta = 0$ in our exact calculation earlier. The $\Delta = 1$ point is different: here we observe the crossover to logarithmic scaling due to the emerging SU(3) symmetry.

\subsubsection*{Process $N_x$}

The staggered magnetization in the $x$ direction is the most straightforward to interpret, as we see fractional behavior with scaling exponents between $0.5$ and $1$, which is in agreement with the expectations. The reason why the exponent is not decreased below $0.58$ is likely the same artifact mentioned above: the interplay between the different subleading terms results in a behavior that looks like a higher-order scaling at short distances. The theoretical connection between the anisotropy parameter and the Hurst exponent is $H=1-\frac{\arccos(-\Delta)}{2\pi}$.

\begin{table}[]
\centering
\begin{tabular}{|l||l|l|l|l|}
\hline
$\Delta$ & $M_z$ & $N_z$ & $M_x$ & $N_x$ \\ \hline \hline
-0.9     & 0 & 0.5   & 0.5   & 0.89  \\ \hline
-0.5     & 0 & 0.5  & 0.5  & 0.82  \\ \hline
0        & 0 & 0.5  & 0.5  & 0.74  \\ \hline
0.5      & 0 & 0.51  & 0.49  & 0.67  \\ \hline
1        & 0 & 0.58  & 0.05  & 0.58  \\ \hline
\end{tabular}
\caption{The numerically fitted Hurst exponent for different magnetizations at selected $\Delta$ parameters.}
\label{tab:hurst}
\end{table}

\begin{table}[]
\centering
\begin{tabular}{r|cc|}
\cline{2-3}
                                      & \multicolumn{1}{c|}{$|\Delta| < 1$}     & $\Delta = 1$     \\ \hline
\multicolumn{1}{|r|}{$M^2_z \propto$} & \multicolumn{2}{c|}{$\log(l)$}                             \\ \hline
\multicolumn{1}{|r|}{$N^2_z \propto$} & \multicolumn{2}{c|}{$l$}                           \\ \hline
\multicolumn{1}{|r|}{$M^2_x \propto$} & \multicolumn{1}{c|}{$l$}                & $ \log(l)$       \\ \hline
\multicolumn{1}{|r|}{\rule{0pt}{1.5em}$N^2_x \propto$} & \multicolumn{2}{c|}{\rule{0pt}{1.5em}$l^{2-\frac{\arccos(-\Delta)}{\pi}}$} \\ \hline
\end{tabular}
\caption{The scaling of the variance for the different magnetizations.}
\label{tab:var_scaling}
\end{table}

\section{Series generation methods}
\label{sec:simulation}

Our quantum chain to classical time series mapping allows us to generate classical series from the ground state (or any other quantum state) of the quantum spin chain. If the quantum system has algebraically decaying long-range correlations and hence non-trivial full counting statistics with self-similarity for its magnetic domains we believe this structure will get transferred to the time series. It is important, however, how series generation can work in practice and how it scales with the length of the time series we want to sample. We discussed the general sampling methodology in an earlier publication \cite{Udvarnoki2025}, but have not focused on the specific complications of fractional series where sampling is known to be a difficult problem classically. 

\subsection{Classical}

For the simulation of fractional Brownian motion, several algorithms have been proposed \cite{Wong2024SimulationApproach, Hosking1984ModelingDifferencing, DAVIES1987TestsEffect}. They all exploit the Gaussian property of the stochastic process and are, therefore, inapplicable to the fractional binomial process. The Davies-Harte is the optimal known exact method \cite{DAVIES1987TestsEffect} that scales as $\mathcal{O}(N\log N)$ with the predefined number of time points ($N$). If the sample size is not known in advance, one can use the Hosking method, which has $\mathcal{O}(N^2)$ scaling.

\subsection{Quantum-inspired}
The exponentially growing state space makes approximating methods necessary to get the ground state of a quantum spin chain longer than a couple of dozen spins. 

One well-established approach uses tensor network states as an ansatz for the ground state of spin chains \cite{Cirac2009RenormalizationLattices}. These try to approximate the high-dimensional quantum state by breaking it up into many interconnected smaller tensors with a specified bond dimension. The simplest type of tensor network states are matrix product states (MPS) \cite {Fannes1992FinitelyChains}, which can be calculated by powerful density-matrix renormalization group (DMRG) algorithms \cite{White1992DensityGroups} or time-evolving block decimation (TEBD) \cite{Vidal2003}, but can only approximate the long-range autocorrelation. Yet, another tensor network ansatz, the multiscale entanglement renormalization ansatz (MERA), was proposed to naturally represent algebraically decaying correlations \cite{Vidal2007EntanglementRenormalization, Evenbly2009AlgorithmsRenormalization}. Both MPS and MERA can be optimized with automatic differentiation-based algorithms \cite{Liao2019DifferentiableNetworks}.

For practical stochastic time-series generation, we would need to sample these ground states. Fortunately, methods already exist for sampling from finite tensor network states \cite{Ferris2012PerfectNetworks, Ferris2012VariationalAnsatz}. We extended these methods for infinite MPS and MERA states, see below.

Sampling from an MPS scales linearly with the number of sites. However, it scales cubically with the bond dimension and the required bond dimension to maintain precision at long distances (or the correlation length) scales with $N^\kappa$, where $\kappa=0.73$ for the Heisenberg model \cite{Tagliacozzo2008ScalingStates}. Sampling from a MERA has a $N^{3\log_3{\chi}}$ scaling, at worst,  but is highly dependent on the contraction order of the tensors. The number of MERA tensors grows only linearly.
In a previous work, we described in detail how an infinite MPS can be sampled \cite{Udvarnoki2025}. Here, we extended these methods to be able to sample from infinite MERA states.

\subsection{Sampling an infinite MPS}

\begin{figure}
    \centering
    \includegraphics[width=1\linewidth]{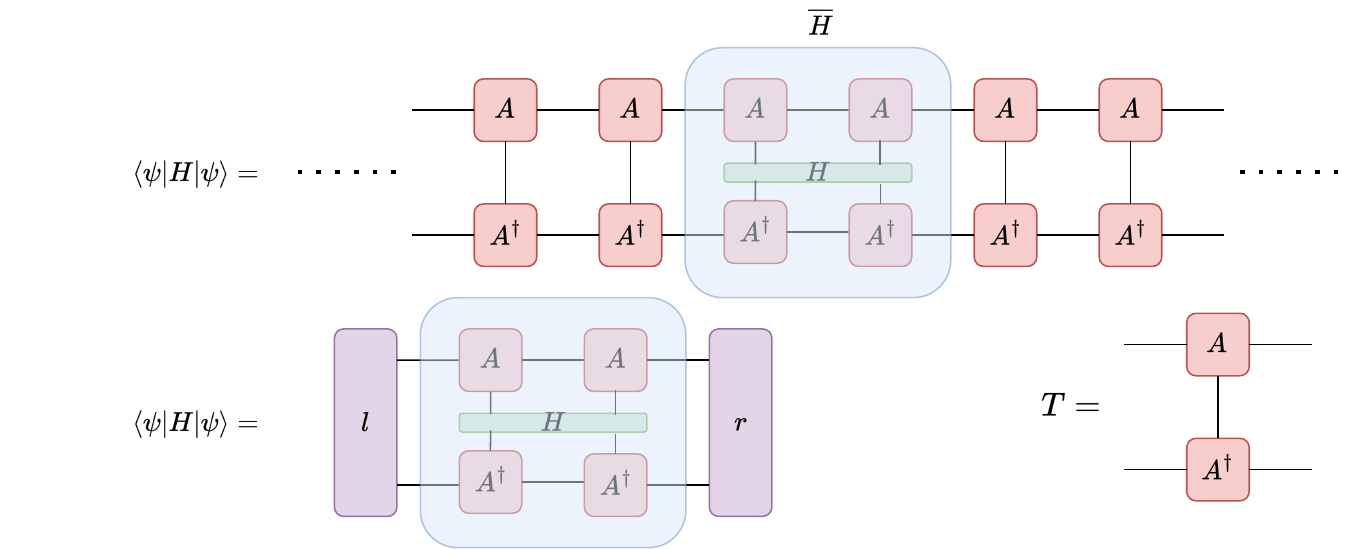}
    \caption{The expectation value of an operator ($H$) can be calculated with the left ($l$) and right ($r$) eigenvectors of the transfer matrix ($T$). The same scheme can be used to calculate the conditional density matrix.}
    \label{fig:MPS}
\end{figure}

\begin{figure}
    \centering
    \includegraphics[width=0.7\linewidth]{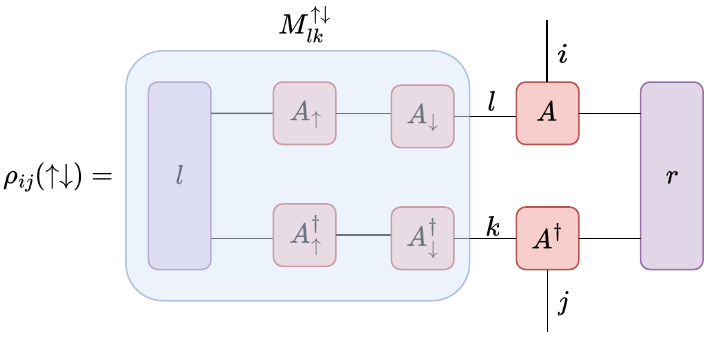}
    \caption{The diagram shows how the three-site conditional density matrix can be calculated, which can be used to sample the third site. $M^{\uparrow \downarrow}_{lk}$ is a two-index tensor that stores all the conditional information.}
    \label{fig:MPS-sampling}
\end{figure}

The stochastic sampling of tensor networks was originally worked out by Ferris and Vidal \cite{Ferris2012PerfectNetworks}. They worked with finite chains and assumed a canonical form such as a unitary tensor network, e.g., uMPS. In case of infinite chains we can relax this criterion but maintain the same linear scaling of the computational cost. We need to calculate the dominant left and right eigenvectors of the MPS transfer matrix ($l$, $r$) and use them as the left and right MPS vectors to close the (very long) chain (see Fig.\ \ref{fig:MPS}) making it effectively infinite. The sampling is done site-by-site. An example for the third site is shown in Fig.\ \ref{fig:MPS-sampling}. The figure shows how the density matrix for the third site is calculated after the first two sites are already fixed in the $\uparrow \downarrow$ states. The density matrix for the third site can be used to sample its spin state with probabilities of up and down:
\begin{equation}
    \begin{aligned}
        P(s_3=\uparrow | s_1=\uparrow, s_2=\downarrow) &= \rho_{\uparrow \uparrow}(\uparrow \downarrow)\\
        P(s_3=\downarrow | s_1=\uparrow, s_2=\downarrow) &= \rho_{\downarrow \downarrow}(\uparrow \downarrow)
    \end{aligned}
\end{equation}
For a detailed description, see our previous work \cite{Udvarnoki2025}. 

\subsection{Sampling an infinite MERA}

Sampling works analogously in other tensor networks. Suppose we have a scale-invariant (infinite) 1D ternary MERA with $T$ transitional layers. The isometries are ($w_0, w_1 \dots w_T, w$), the disentanglers ($u_0, u_1 \dots u_T, u$).
We can calculate the average two-site density matrix ($\rho_2$) of the scale-invariant layers by finding the dominant eigenoperator of the average descending superoperator. 

\subsubsection{The average three-site MERA density matrix}
\begin{figure}
    \centering
    \includegraphics[width=1\linewidth]{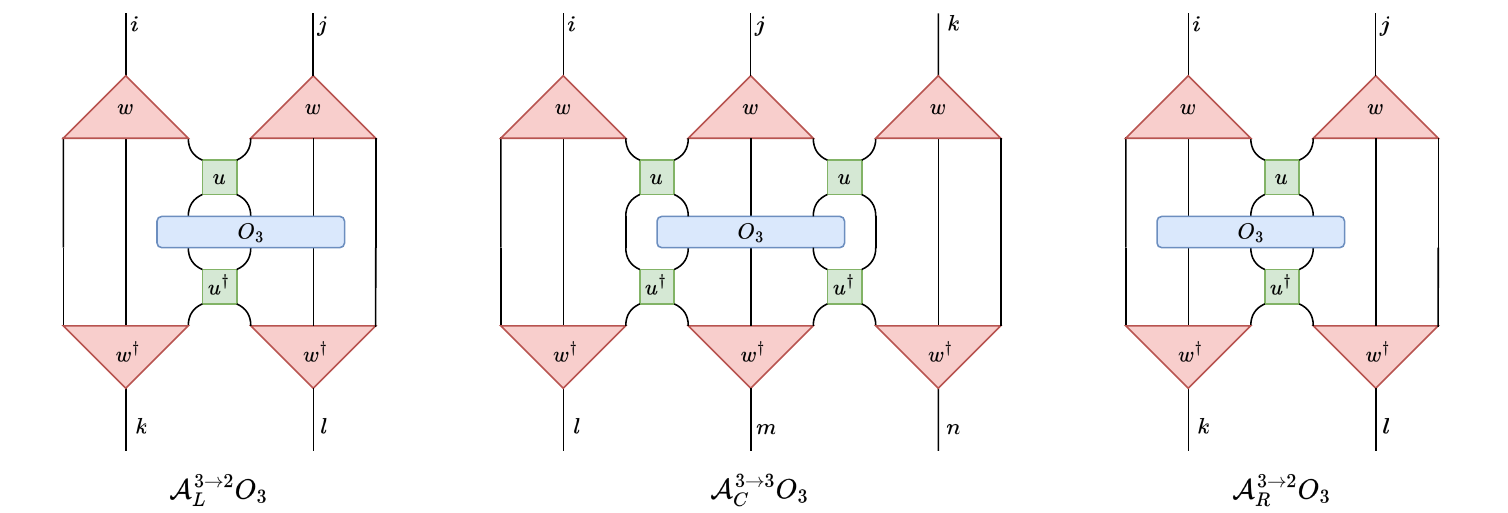}
    \caption{A three-site operator can be lifted to the next layer in three different ways based on its relative positive in the network. This is represented by the three ascending superoperators $\mathcal{A}^{3\rightarrow2}_L$ (left), $\mathcal{A}^{3\rightarrow3}_C$ (middle), $\mathcal{A}^{3\rightarrow2}_R$ (right).  $\mathcal{A}^{3\rightarrow2}_L$ and $\mathcal{A}^{3\rightarrow2}_R$ maps the three-site operator to a two-site operator, while $\mathcal{A}^{3\rightarrow3}_C$ maps it to a three-site operator in the next layer.}
    \label{fig:MERA_ascending_3}
\end{figure}

In a ternary MERA, every $N$-site operator will eventually be mapped to a two- or three-site operator after enough layers. Three-site operators are, however, mapped to two- or three-site operators in the next layer, depending on their position, making infinitely many layers necessary to map them to a two-site operator. This is done by the three three-site ascending superoperators $\mathcal{A}^{3\rightarrow2}_L$, $\mathcal{A}^{3\rightarrow3}_C$, $\mathcal{A}^{3\rightarrow2}_R$, shown in Fig. \ref{fig:MERA_ascending_3}. They all occur on $\frac{1}{3}$th of the positions in every layer. 

We can derive an analytical expression for the average three-site density matrix to allow sampling and the calculation of expectation values for all $N$-site operators.

Suppose we calculate the expectation value of a three-site operator $O_3$ - according to the middle-ground contraction scheme - by applying the ascending superoperator on $O_3$ until the three-site operator is mapped to a two-site operator and applying the descending superoperator on the average two-site density matrix $\bar{\rho}_2$ to descend the density matrix to the same layer:

\begin{widetext}
\begin{equation}
\label{eq:3expectation}
\begin{split}
\langle O_3 \rangle &= \lim_{n\rightarrow\infty}\mathrm{Tr}\Bigg[\mathcal{D}^{n-1}\bar{\rho}_2 \frac{\mathcal{A}^{3\rightarrow2}_L +\mathcal{A}^{3\rightarrow2}_R}{3} O_3 +  \mathcal{D}^{n-2}\bar{\rho}_2 \frac{\mathcal{A}^{3\rightarrow2}_L + \mathcal{A}^{3\rightarrow2}_R}{3^2} \mathcal{A}^{3\rightarrow3}_C O_3 +\dots  \\ 
& \qquad \qquad \quad+ \mathcal{D}^0 \bar{\rho}_2 \frac{\mathcal{A}^{3\rightarrow2}_L +\mathcal{A}^{3\rightarrow2}_R}{3^n} \left(\mathcal{A}^{3\rightarrow3}_C\right)^{n-1} O_3\Bigg] \\
&=\mathrm{Tr}\left[ \bar{\rho}_2 \frac{\mathcal{A}^{3\rightarrow2}_L +\mathcal{A}^{3\rightarrow2}_R}{3} \sum_{n=0}^{\infty} \frac{\left(\mathcal{A}^{3\rightarrow3}_C\right)^{n}}{3^n}   O_3    \right] \\
&=\mathrm{Tr}\left[ \bar{\rho}_2 \frac{\mathcal{A}^{3\rightarrow2}_L +\mathcal{A}^{3\rightarrow2}_R}{3} \left(I-\frac{\mathcal{A}^{3\rightarrow3}_C}{3} \right)^{-1} O_3    \right],
\end{split}
\end{equation}
\end{widetext}
where we exploited the fact that $\bar{\rho}_2$ is an eigenoperator of $\mathcal{D}$ with eigenvalue $1$. The condition for closed-form substitution of the infinite sum in the last line is that the dominant eigenvalue of $\mathcal{A}^{3\rightarrow3}_C$ is smaller in absolute value than three.

If we define the average three-site ascending superoperator $\mathcal{A}^{3\rightarrow2}$:
\begin{equation}
\label{eq:3ascending}
    \mathcal{A}^{3\rightarrow2} = \frac{\mathcal{A}^{3\rightarrow2}_L +\mathcal{A}^{3\rightarrow2}_R}{3} \left(I-\frac{\mathcal{A}^{3\rightarrow3}_C}{3} \right)^{-1}
\end{equation}
and consider that:
\begin{equation}
    \langle O_3 \rangle = \mathrm{Tr}\left(\bar{\rho}_3 O_3\right) = \mathrm{Tr}\left(\bar{\rho}_2 \mathcal{A}^{3\rightarrow2} O_3\right)
\end{equation}
is true for arbitrary $O_3$. We see that the average three-site density matrix $\bar{\rho}_3$ is:
\begin{equation}
    \bar{\rho}_3 = \bar{\rho}_2 \mathcal{A}^{3\rightarrow2}
\end{equation}

Calculating the inverse in \eqref{eq:3ascending} can be an obstacle due to the $\chi^{12}$ size of $\mathcal{A}^{3\rightarrow3}_C$. In this case, one can fall back to the infinite sum form of \eqref{eq:3expectation} and calculate it approximately.

\subsubsection{Sampling}

If we want to sample $N$ sites from MERA, we build a MERA network with $N$ physical sites and with at least $T$ layers, but as many as we need to coarse-grain them to two or three sites and then connect them to the average two- or three-site density matrix. 

The resulting network is an $N$-site density matrix with $2N$ free indices. Of course, this would be untreatable in the matrix form. For the sequential sampling from left to right, we first trace out the rightmost $2N-2$ indices and take a sample according to the probabilities of the resulting one-site density matrix. In the next step, we fix the first site, trace out $2N-4$ indices, and take a sample according to the resulting one-site density matrix. We repeat the steps until no free indices are left, and then we have a length $N$ sample from the MERA network.

\begin{figure}
    \centering
    \includegraphics[width=0.7\linewidth]{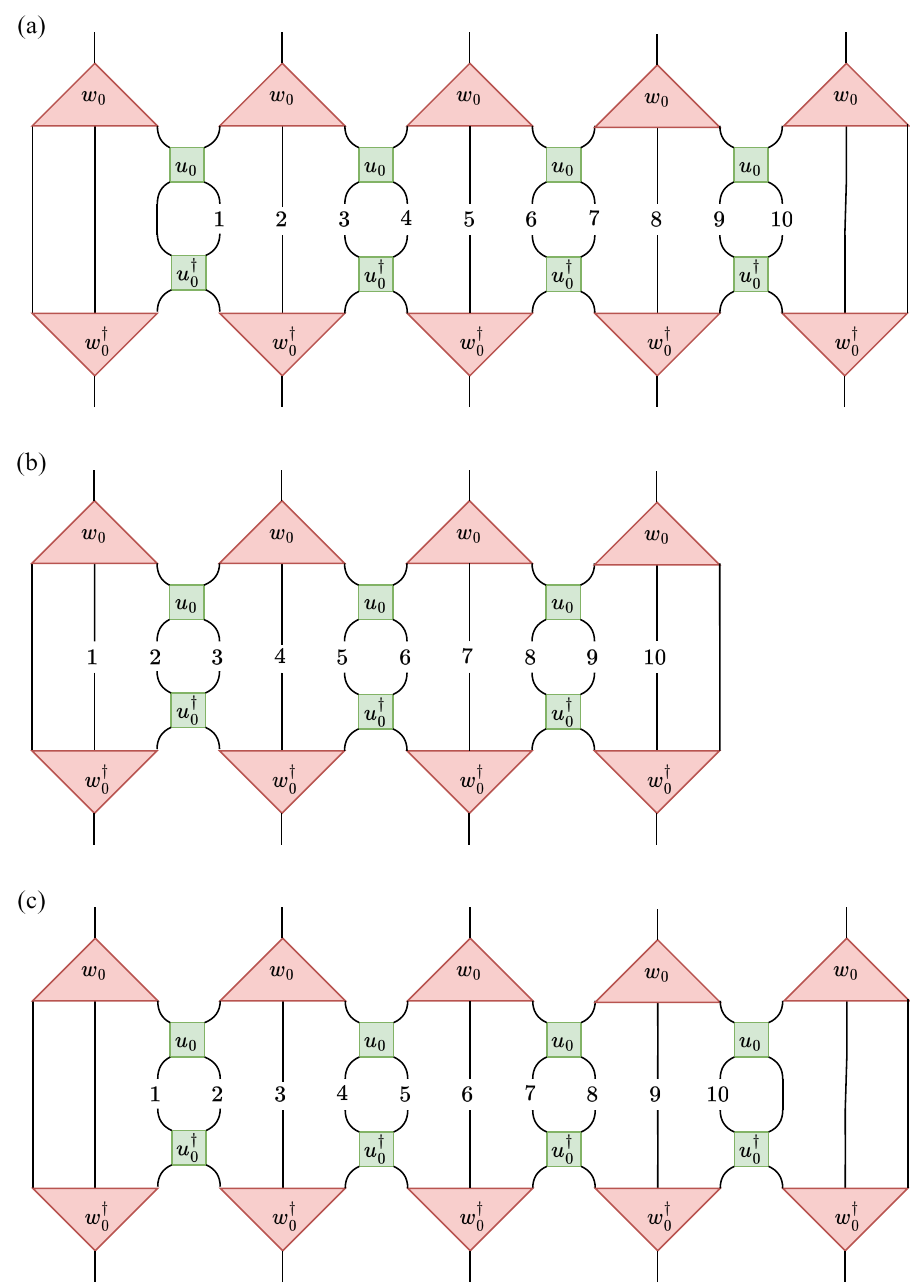}
    \caption{The three different layouts of a 10-site operator in the first layer. Note that the same freedom exists in the second layer, which is not represented here.}
    \label{fig:MERA_N}
\end{figure}

\begin{figure}
    \centering
    \includegraphics[width=0.7\linewidth]{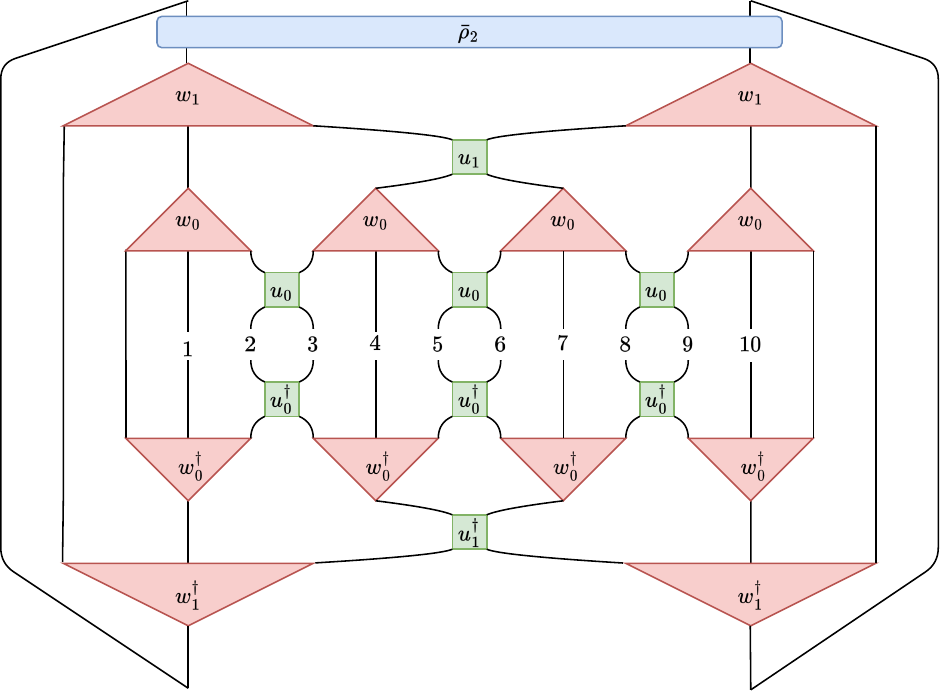}
    \caption{An example ternary MERA network with $T=2$ transitional layers for sampling $N=10$ sites. The one-site density matrix of the next site is calculated by fixing the already sampled indices, leaving one pair free, and tracing out the rest of them.}
    \label{fig:MERA_net_c}
\end{figure}

For an exact result, one would need to average all possible layouts, in other words, over every different $N$-site density operator. As there are three possible layouts per layer, as shown in Fig. \ref{fig:MERA_N}, this multiplies the computational cost by three with every renormalization layer. However, it was previously shown \cite{Evenbly2013QuantumAnsatz} that fluctuations due to the violation of translational invariance of the MERA network are of order $10^{-4}$ and diminish with growing bond dimension. For the sake of simplicity and demonstration, we used only a single layout, where the first free index is the middle leg. Such a MERA network for sampling $N=10$ sites is shown in Fig. \ref{fig:MERA_net_c}.

Calculating expectation values of two-site operators like two-point correlation can be done in a very similar way, e.g., with the network in Fig. \ref{fig:MERA_net_c}. We just put the two operators on the particular sites and trace the rest.

\begin{figure}
    \centering
    \includegraphics[width=1\linewidth]{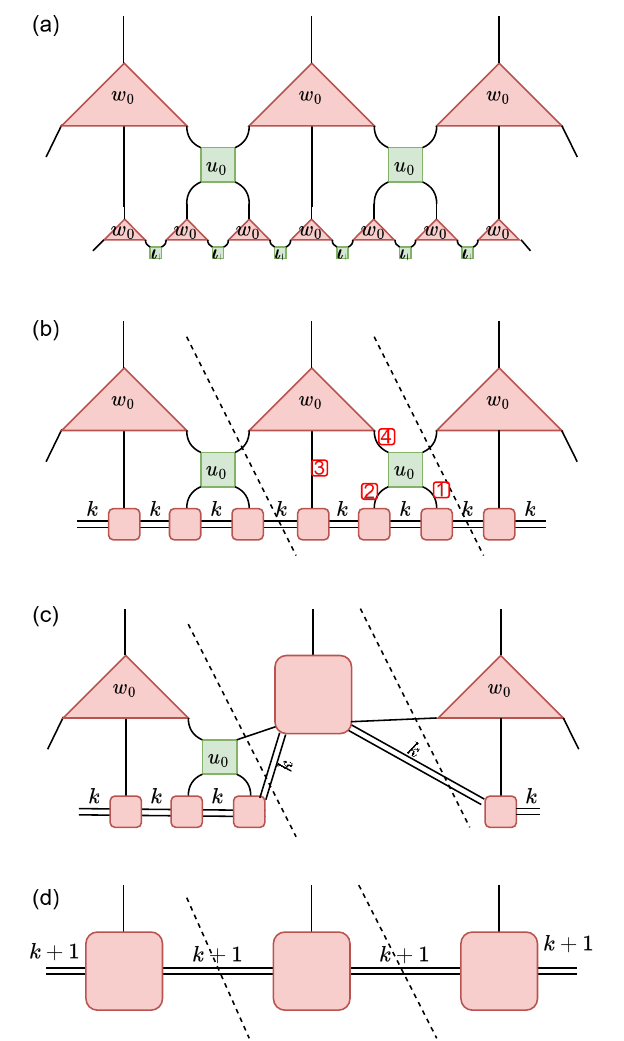}
    \caption{A simple systematic contraction order of a scale-invariant MERA network. We start at the top figure and move down by contracting the tensors. The dashed line encloses one unit. Red numbers label the contraction order of legs. Double lines with a label $k$ denote that the tensors have $k$ common indices.}
    \label{fig:contraction}
\end{figure}

Finding the optimal contraction order is a hard task. However, the computational cost can be estimated with a systematic contraction order. Such a scheme is shown in Fig. \ref{fig:contraction}. Suppose that the $n-1$ sites are already fixed and we want to sample the $n$th site. We start at the bottom layer in the middle of the network (Fig. \ref{fig:contraction} (a)), where no free indices are present. First, we contract the isometries with the disentanglers in the bottom layer. This is a $\chi^4$ operation. In Fig. \ref{fig:contraction} (b), we see the resulting network, with the additional generalization that the tensors on the bottom are connected with $k$ parallel bonds. This is because, as we will see, moving up layer by layer, only the effective bond dimension of this line will multiply. To see this, we contract all the tensors between the dashed lines to end up with the network in Fig.\ \ref{fig:contraction}(c).  This can be done with cost scaling as $\chi^{3k+4}$ in the order shown in the figure. After doing it, we end up with one less layer, and the new bottom layer will have a form similar to the previous one, only the tensors at the bottom will have one more parallel bond between them. 

If we are contracting the part with the free index, that will only multiply the cost by the physical dimension. For the whole network with $\log_3(n)$ layers and $\frac{n}{3^k}$ tensors in the $k$th layer, the computational cost scales with:
\begin{equation}
\sum_{k=1}^{\log_3{n}} \chi^{3k+4} \frac{n}{3^k} = \frac{\chi^7\left(\chi^{3\log_3(n)} - n\right)}{(\chi^3-3)} \propto n ^{3\log_3{\chi}} + O(n).
\end{equation}

To obtain an $N$-site sample, one would need to do the above contraction for $n=1,2 \dots N$. This leads to a computational cost that is upper bounded by $N ^{ 3\log_3{\chi} +1}$.

\section{Conclusions}
\label{sec:conclusion}
Power-law scaling is a characteristic feature of both one-dimensional quantum systems at their critical point and classical fractional time series. This formal similarity can be made rigorous, and a mapping can be established that defines classical fractional series from quantum systems and creates a means to generate/sample these series in a sequential way. The concepts of this mapping were described in an earlier publication \cite{Udvarnoki2025}. In this paper we focused on the mapping between the ground state of the spin-1/2 Heisenberg XXZ chain and four different fractional stochastic processes of two states built upon this quantum model.

The Heisenberg model has an extended critical region in which critical exponents depend on a parameter called anisotropy. This brings the hope that these two-state fractional processes can be created with different Hurst exponents. A priori, simply looking at the asymptotic decay of quantum correlations, we expected to see Hurst exponents above and below $1/2$. Our analysis substantiated this expectation for $H\ge 1/2$. It is easy to generate fractional processes with positive persistence from the spin chain. On the other hand, we identified a difficulty for $H<1/2$, i.e., anti-persistent (rough) behavior. We realized that our mapping directly introduces a scaling operator with linear scaling, which in itself leads to $H=1/2$. When this operator is present the anti-persistent behavior is suppressed and non-fractional scaling prevails. There is only one case when this operator vanishes: continuous quantum symmetries dictate a special sum rule of the quantum correlation function, and as a result, the linear operator is canceled. A $U(1)$ symmetry is present in one of the processes we defined for any value of the chain anisotropy, and an $SU(2)$ symmetry in a special point of the parameter space for two of the processes. In these cases we do see $H<1/2$. However, in the XXZ Heisenberg chain this can only lead to $H=0$ exactly with logarithmic scaling. It is an open question if there is any other one-dimensional quantum system where a quantum symmetry kills the dangerous linear operator, but it still possesses a scaling operator in its spectrum with $H>0$. This question is left for future work.

In the second half of the paper, we discussed how fractional series can be generated from a tensor network representation of the quantum state. The simplest generation method works with MPS (matrix product) states \cite{Schollwock2011TheStates}. This has linear scaling in the chain length. Note, however, that the MPS is only an approximation of the quantum state, and its representation accuracy becomes worse close to the critical point. The MPS requires more and more parameters to keep accuracy as the critical point is approached, and this ruins linear scaling at the critical point. Other tensor networks have been proposed in the literature for better approximating quantum critical systems, such as MERA.\cite{Vidal2007EntanglementRenormalization, Evenbly2009AlgorithmsRenormalization} Here we extended sampling from MERA states for infinite chains and discussed how MERA can be used to generate a classical fractional process in this context. Working with MERA, though, has its own complexity, and it is unclear if the complexity of the process is counterbalanced by the smaller number of parameters MERA requires for the same precision. Again, this is left for future studies.

\section{Acknowledgments}
This work was supported by the Ministry of Innovation and Technology and the National Research, Development and Innovation Office of Hungary (NKFIH) within the National Quantum Technology Program (Grant No. 2022-2.1.1-
NL-2022-00004) and Grant Nos.~K134983, TKP2021-NVA-04. 
\"O.L. acknowledges support of the Hans Fischer Senior Fellowship programme funded by the Technical
University of Munich - Institute for Advanced Study, and of the Center
for Scalable and Predictive methods
for Excitation and Correlated phenomena (SPEC), funded as part of the
Computational Chemical Sciences Program FWP 70942 by the U.S. Department of
Energy (DOE), Office of Science, Office of Basic Energy Sciences, Division
of Chemical Sciences, Geosciences, and Biosciences at Pacific Northwest
National Laboratory.
M.A.W. has also been supported by the Janos Bolyai Research Scholarship of the Hungarian Academy of Sciences.
Z.U. acknowledges the support of the Doctoral Student Scholarship Program of the Co-operative Doctoral Program of the Ministry of Innovation and Technology, financed by the National Research, Development and Innovation Fund of Hungary.

\appendix

\subsection{Auxiliary functions}

We list first some special functions that will be used in the following derivations. The Riemann zeta function $\zeta(a)$ and the Dirichlet eta function $\eta(a)$ are defined as
\begin{equation}
    \zeta(a) = \sum_{k=1}^\infty \frac{1}{k^a}, \qquad
    \eta(a) = \sum_{k=1}^\infty \frac{(-1)^{k-1}}{k^a}.
\end{equation}
We will need the partial sums
\begin{equation}
    H_n^{(a)} = \sum_{k=1}^n \frac{1}{k^a}, \qquad
    D_n^{(a)} = \sum_{k=1}^n \frac{(-1)^{k-1}}{k^a}.
\end{equation}
where $H_n^{(a)}$ is called the {\it generalized harmonic number}. Easy to see that the two are connected; for $n$ even, we get
\begin{equation}
    D_n^{(a)} = H_n^{(a)} - 2^{1-a} H_{n/2}^{(a)}.
\end{equation}
which can be compared to the infinite sum result
\begin{equation}
    \eta^{(a)} = \left(1 - 2^{1-a}\right) \zeta^{(a)}.
\end{equation}

For positive integer powers $m \in \mathcal{N}$, these sums can also be expressed using the polygamma functions \cite{Kolbig1996TheX=34}:
\begin{eqnarray}
    H_n &\equiv& \sum_{k=1}^n \frac{1}{k} = \gamma-\psi^{(0)}(n+1) \\
    H_n^{(m)} &\equiv& \sum_{k=1}^n \frac{1}{k^{m}} \\
    &=& \zeta(m)-\frac{\psi^{(m-1)}(n+1)}{(-1)^{(m)}(m-1)!}, \quad m\ge 2 
\end{eqnarray}
where $\gamma=0.5772...$ is the Euler-Mascheroni constant, and $\psi^{(m)}$ is the $m$-th derivative of the digamma function (aka polygamma). In particular for $m=2$ and $n$ even:
\begin{eqnarray*}
    &&H_n^{(2)} \equiv \sum_{k=1}^n \frac{1}{k^2} = \zeta(2)-\psi^{(1)}(n+1) =
    \frac{\pi^2}{6}-\psi^{(1)}(n+1) \\
    && D_n^{(2)} \equiv \sum_{k=1}^n \frac{(-1)^{k-1}}{k^2} =
    \frac{\pi^2}{12}-\psi^{(1)}(n+1) + \frac{1}{2}\psi^{(1)}\left(\frac{n}{2}+1\right)
\end{eqnarray*}
Later, we will need for $n$ even
\begin{eqnarray*}
    \sum_{k=1}^{n/2} \frac{1}{(2k-1)^2} &=& \frac{1}{2}\left(H_n^{(2)} + D_n^{(2)}\right) \\&=& \frac{\pi^2}{8}-\psi^{(1)}(n+1) + \frac{1}{4}\psi^{(1)}\left( \frac{n}{2}+1 \right) \\
    &=& \frac{\pi^2}{8}-\frac{1}{4}\psi^{(1)}\left(\frac{n}{2}+\frac{1}{2}\right)
\end{eqnarray*}
where for the last equality see https://en.wikipedia.org/wiki/Trigamma\_function.

For large $n$ the Euler-Maclaurin \cite{concrete} sum formula can be used:
\begin{eqnarray*}
    H_n^{(a)} &=& \sum_{k=1}^n \frac{1}{k^a}\\ &=& \zeta(a) -\frac{1}{(a-1)n^{a-1}} +\frac{1}{2n^{a}} - \frac{a}{12 n^{a+1}} \\&&+ O\left(\frac{1}{n^{a+2}}\right), \quad \rm{if}\; a > 1, \\
    H_n &=& \sum_{k=1}^n \frac{1}{k} \\&=& \log(n) +\gamma +\frac{1}{2n} -\frac{1}{12 n^2} + O\left(\frac{1}{n^3}\right),
    \quad \rm{if}\; a = 1,
\end{eqnarray*}
and for the Dirichlet sum for $n$ even, we find
\begin{eqnarray*}
    D_n^{(a)} &=& \sum_{k=1}^n \frac{(-)^{k-1}}{k^a} \\&=& \left(1-\frac{1}{2^{a-1}}\right)\zeta(a) -\frac{1}{2n^{a}} \\ &&- \frac{a}{4 n^{a+1}} + O\left(\frac{1}{n^{a+2}}\right), \quad \rm{if}\; a > 1, \\
    D_n &=& \sum_{k=1}^n \frac{(-)^{k-1}}{k} \\&=& \log(2) -\frac{1}{2n} +\frac{1}{4 n^2} + O\left(\frac{1}{n^3}\right),
    \quad \rm{if}\; a = 1,
\end{eqnarray*}
The first we found empirically also valid for $a>0$ (using the analytic continuation of the zeta function for $a<1$). For $n$ odd we have a slightly different expansion.

\section{Alternatig power sum}
\label{appendix:altsum}
\subsection{Single sum}
We can write the alternating partial sum as the difference between two non-alternating partial sums:
\begin{equation}
\sum_{k=1}^n (-1)^k k^p = -D_l^{(-p)} = 
\begin{cases}
2^{p+1} H^{(\text{-}p)}_{\frac{n}{2}} - H_n^{(\text{-}p)} & \text{n even}\\
\\
2^{p+1} H^{(\text{-}p)}_{\frac{n-1}{2}} - H_n^{(\text{-}p)} & \text{n odd}
\end{cases}
\end{equation}

We will use the series expansion of the generalized harmonic numbers $H_n^p$ (obtained with the help of the Euler-McLaurin formula) to obtain the asymptotic behavior of the sum. It is valid for any real $p\neq-1$.
\begin{equation}
    H_n^{(\text{-}p)} = \frac{n^{p+1}}{p+1} + \frac{n^p}{2} + \frac{pn^{p-1}}{12} + O(n^{p-3}) + \zeta(-p),
\end{equation}
where $\zeta$ denotes the Riemann zeta function.

For $p=-1$ the behavior is similar:
\begin{equation}
    H_n^{(\text{-}1)} = \ln(n) + \frac{n^p}{2} + \frac{pn^{p-1}}{12} + O(n^{p-3}) + \gamma,
\end{equation}
where $\gamma$ is the Euler-Mascheroni constant.

For even $l$:
\begin{equation}
\label{eq:s_even}
-D_n^{(-p)} = \frac{n^p}{2} + \frac{pn^{p-1}}{4} + C_1 + O(n^{p-3})
\end{equation}

For odd $l$:
\begin{align}
\label{eq:s_odd}
-D_n^{(-p)} &= 2^{p+1}H^{(\text{-}p)}_{\frac{n-1}{2}}- H_{n-1}^{(\text{-}p)} - n^p \nonumber \\
&= \frac{(n-1)^p}{2} -n^p + \frac{p(n-1)^{p-1}}{4} +  C_1 +O(n^{p-3}) \nonumber \\
&= -\frac{n^p}{2} - \frac{pn^{p-1}}{4} + C_1+ O(n^{p-3}) 
\end{align}

Now if combine \eqref{eq:s_even} and \eqref{eq:s_odd} we end up with the general formula:
\begin{equation}
\label{eq:single_alternating_sum}
    \sum_{k=1}^n (-1)^k k^p = \frac{(-1)^n}{2}\left[n^p + \frac{pn^{p-1}}{2} \right] + C_1 + O(n^{p-3})
\end{equation}

\begin{equation}
C_1 =
\begin{cases}
(2^{p+1}-1)\zeta(-p) & \text{for}\; p \neq -1\\
\\
\ln(2) & \text{for} \; p=-1
\end{cases}
\end{equation}

\subsection{Double sum}
\begin{equation}
    \sum_{n=1}^l \sum_{k=1}^n(-1)^kk^p = \sum_{n=1}^l -D_n^{(-p)}
\end{equation}

\begin{align}
    \sum_{n=1}^l-D_n^{(-p)} &= \frac{\sum_{n=1}^l(-1)^nn^p}{2} + \frac{p\sum_{n=1}^l(-1)^nn^{p-1}}{4} \nonumber \\
    & \quad + \sum_{n=1}^l C_1 + \sum_{n=1}^lO(n^{p-3}) \nonumber \\
    &= -D_l^{(-p)} -D_n^{(-p+1)} + C_1 l \nonumber \\
    & \quad + O(l^{p-2}) + const. \nonumber \\
    &=\frac{(-1)^l}{4}\left[l^p + p l^{p-1}\right] \nonumber \\
    & \quad + C_1 l + const. + O(l^{p-2}),
\end{align}
 for every $p\neq2$.

\section{Exact calculations of magnetization variance ($\Delta = 0$)}
\label{sec:exact_var}

At the XX point, $\Delta=0$, the exact form of the autocovariance function is known for the Heisenberg spin chain \cite{Kitanine2002CorrelationRepresentations}, see Eqs.\ (\ref{eq:autocov_z}) and (\ref{eq:autocov_x}). This makes it possible to calculate the variance of the magnetization without any uncertainty.

\subsection{Uniform magnetization in $z$ direction ($M_z$)}
Introduce
\begin{equation}
V_l = \langle M^2(l) \rangle = \frac{1}{4}l + 2 \sum_{i=1}^{l-1} \sum_{j=i+1}^l C(j-i)
\end{equation}
and the difference
\begin{equation}
\Delta V_{l+1} = V_{l+1}- V_l = \frac{1}{4} + 2 \sum_{m=1}^{l} C(m)
\end{equation}

Substituting in the autocovariance function:
\begin{eqnarray}
\Delta V_{l+1}^z &=& \frac{1}{4} + 2 \sum_{k=1}^{l} C^z(k) 
= \frac{1}{4} + 2 \sum_{k=1}^{l} \frac{1}{2\pi^2 k^2} \left[(-)^k -1\right] \nonumber \\
&=& \frac{1}{4} - \frac{1}{\pi^2} \left[H_l^{(2)}+D_l^{(2)}\right]
\nonumber \\ &=& \frac{1}{4} - \frac{1}{\pi^2} \left[\frac{3}{2}\zeta(2) - \frac{1}{l} -\frac{1}{3l^3} + O\left(\frac{1}{l^4}\right)\right]
\end{eqnarray}
Using $\zeta(2)=\pi^2/6$, the constant term cancels, and we get
\begin{equation}
\Delta V_{l+1}^z = \frac{1}{\pi^2 l} +\frac{1}{3\pi^2 l^3} + O\left(\frac{1}{l^4}\right)
\end{equation} 
Or alternatively, with the exact form above
\begin{eqnarray}
\Delta V_{l+1}^z &=& \frac{1}{4} - \frac{1}{\pi^2} \left[H_l^{(2)}+D_l^{(2)}\right] \nonumber \\
&=& 
\frac{1}{4} - \frac{1}{\pi^2} \left[ \frac{1}{4}\left(\pi^2-2\psi^{(1)}(l/2+1/2)\right) \right] \nonumber \\
&=& \frac{\psi^{(1)}(l/2+1/2)}{2\pi^2} \nonumber \\ &=& \frac{1}{\pi^2 l} +\frac{1}{3\pi^2 l^3} +\frac{7}{15\pi^2 l^5} + O\left(\frac{1}{l^6}\right)
\end{eqnarray} 
With this, the variance of the uniform domain magnetization in the $z$ direction reads
\begin{eqnarray}
V_l^z &=& \sum_{l^\prime=1}^{l} \Delta V_{l^\prime}^z = \frac{1}{2\pi^2} \sum_{l^\prime=0}^{l-1}  \psi^{(1)}(l^\prime /2+1/2) \nonumber \\ &=& \frac{1}{2\pi^2} \left[ \sum_{l^\prime=1}^{l/2}  \psi^{(1)}(l^\prime) + \sum_{l^\prime=0}^{l/2-1}  \psi^{(1)}(l^\prime+1/2) \right]
\end{eqnarray} 
The sums on the right-hand side have closed-form results as
\begin{eqnarray}
\sum_{l^\prime=1}^{l/2}  \psi^{(1)}(l^\prime) &=& \psi^{(0)}\left(\frac{l}{2}+1\right) +\frac{l}{2} \psi^{(0)}\left(\frac{l}{2}+1\right)+\gamma \nonumber \\
\sum_{l^\prime=0}^{l/2-1}  \psi^{(1)}(l^\prime+1/2) &=& \psi^{(0)}\left(\frac{l}{2}+\frac{1}{2}\right) +\frac{l-1}{2} \psi^{(0)}\left(\frac{l}{2}+\frac{1}{2}\right) \nonumber \\ &&+\frac{\pi^2}{4}+\ln 4+\gamma,
\end{eqnarray} 
which gives
\begin{eqnarray}
V_l^z &=& \frac{1}{2\pi^2} \Bigl[ \psi^{(0)}\left(\frac{l}{2}+1\right) +\frac{l}{2} \psi^{(0)}\left(\frac{l}{2}+1\right) \nonumber \\
&&+\psi^{(1)}\left(\frac{l}{2}+\frac{1}{2}\right) + 
\frac{l-1}{2} \psi^{(1)}\left(\frac{l}{2}+\frac{1}{2}\right) \nonumber \\ 
&& +\frac{\pi^2}{4}+\ln 4+2\gamma \Bigr] \nonumber \\
&=& 2\ln l + \left[\frac{\pi^2}{4}+2(\gamma+1)\right] -\frac{1}{l}+\frac{6}{l^2}
\end{eqnarray} 
As we see, it scales sublinearly. Note that
\begin{equation}
    \frac{\pi^2}{4}+2(\gamma+1) = 5.621832430.
\end{equation}

\subsection{Staggered magnetization in $z$ direction ($N_z$)}
Introduce
\begin{equation}
W_l = \langle N^2(l) \rangle = \frac{1}{4}l + 2 \sum_{i=1}^{l-1} \sum_{j=i+1}^l (-)^{j-i} C(j-i)
\end{equation}
and the difference
\begin{equation}
\Delta W_{l+1} = W_{l+1}- W_l = \frac{1}{4} + 2 \sum_{m=1}^{l} (-)^m C(m)
\end{equation}

Substituting in the autocovariance function:
\begin{eqnarray*}
\Delta W_{l+1}^z &=& \frac{1}{4} + 2 \sum_{k=1}^{l} (-)^k C^z(k) \nonumber \\
&=& \frac{1}{4} + 2 \sum_{k=1}^{l} \frac{1}{2\pi^2 k^2} (-)^k \left[(-)^k -1\right] \nonumber \\
&=& = \frac{1}{4} - 2 \sum_{k=1}^{l} \frac{1}{2\pi^2 k^2}  \left[(-)^k -1\right] \nonumber  \\
&=& \frac{1}{4} + \frac{1}{\pi^2} \left[H_l^{(2)}+D_l^{(2)}\right] \nonumber \\
&=& \frac{1}{4} + \frac{1}{\pi^2} \left[\frac{3}{2}\zeta(2) - \frac{1}{l} -\frac{1}{3l^3} + O\left(\frac{1}{l^4}\right)\right]
\end{eqnarray*}
Using $\zeta(2)=\pi^2/6$, now the constant term does not cancel, and we get
\begin{equation}
\Delta W_{l+1}^z = \frac{1}{2} -\frac{1}{\pi^2 l} -\frac{1}{3\pi^2 l^3} + O\left(\frac{1}{l^4}\right)
\end{equation} 
Or alternatively, with the exact form above
\begin{eqnarray*}
\Delta W_{l+1}^z &=& \frac{1}{4} + \frac{1}{\pi^2} \left[H_l^{(2)}+D_l^{(2)}\right] \nonumber \\
&=& 
\frac{1}{4} + \frac{1}{\pi^2} \left[ \frac{1}{4}\left(\pi^2-2\psi^{(1)}(l/2+1/2)\right) \right] \nonumber \\
&=& \frac{1}{2} -\frac{\psi^{(1)}(l/2+1/2)}{2\pi^2} \nonumber \\
&=& \frac{1}{2} -\frac{1}{\pi^2 l} -\frac{1}{3\pi^2 l^3} -\frac{7}{15\pi^2 l^5} + O\left(\frac{1}{l^6}\right)
\end{eqnarray*} 
With this, the variance of the staggered domain magnetization in the $z$ direction reads
\begin{eqnarray*}
W_l^z &=& \sum_{l^\prime=1}^{l} \Delta W_{l^\prime}^z = \frac{l}{2} -\frac{1}{2\pi^2} \sum_{l^\prime=0}^{l-1}  \psi^{(1)}(l^\prime /2+1/2) \nonumber \\ &=& \frac{l}{2} -\frac{1}{2\pi^2} \left[ \sum_{l^\prime=1}^{l/2}  \psi^{(1)}(l^\prime) + \sum_{l^\prime=0}^{l/2-1}  \psi^{(1)}(l^\prime+1/2) \right]
\end{eqnarray*} 
Again, the sums on the right-hand side have the closed-form results
\begin{eqnarray}
\sum_{l^\prime=1}^{l/2}  \psi^{(1)}(l^\prime) &=& \psi^{(0)}\left(\frac{l}{2}+1\right) +\frac{l}{2} \psi^{(0)}\left(\frac{l}{2}+1\right)+\gamma \nonumber \\
\sum_{l^\prime=0}^{l/2-1}  \psi^{(1)}(l^\prime+1/2) &=& \psi^{(0)}\left(\frac{l}{2}+\frac{1}{2}\right) +\frac{l-1}{2} \psi^{(0)}\left(\frac{l}{2}+\frac{1}{2}\right) \nonumber \\ && +\frac{\pi^2}{4}+\ln 4+\gamma,
\end{eqnarray} 
which gives
\begin{eqnarray}
W_l^z &=& \frac{l}{2} -\frac{1}{2\pi^2} \Bigl[ \psi^{(0)}\left(\frac{l}{2}+1\right) +\frac{l}{2} \psi^{(0)}\left(\frac{l}{2}+1\right) \nonumber \\ 
&&+\psi^{(1)}\left(\frac{l}{2}+\frac{1}{2}\right) + 
\frac{l-1}{2} \psi^{(1)}\left(\frac{l}{2}+\frac{1}{2}\right) \nonumber \\
&&+\frac{\pi^2}{4}+\ln 4+2\gamma \Bigr]  \nonumber \\
&=& \frac{l}{2} -2\ln l - \left[\frac{\pi^2}{4}+2(\gamma+1)\right] +\frac{1}{l} -\frac{6}{l^2}
\end{eqnarray} 
As we see, it scales linearly with a log correction as the subleading term. So the Hurst exponent is $H=1/2$.

\clearpage
\bibliography{main}

\end{document}